# Lower Critical Field, Anisotropy, and Two-Gap Features of LiFeAs

K. Sasmal,[1] B. Lv,[2] Z. Tang,[2] F. Y. Wei,[1] Y. Y. Xue,[1] A. M. Guloy,[2] and C. W. Chu[1,3,4]

[1]Department of Physics and TCSUH, University of Houston, Houston, Texas 77204-5002, USA

[2]Department of Chemistry and TCSUH, University of Houston, Houston, Texas 77204-5003, USA

[3]Lawrence Berkeley National Laboratory, 1 Cyclotron Road, Berkeley, California 94720, USA

[4]Hong Kong University of Science and Technology, Hong Kong, China

**Abstract**

The magnetic properties of LiFeAs, as single crystalline and polycrystalline samples, were investigated. The lower critical field deduced from the vortex penetration of two single crystals appears to be almost isotropic with a temperature dependence closer to that of two-gap superconductors. The parameters extracted from the reversible magnetizations of sintered polycrystalline samples are in good agreement with those from the single crystal data.

The discovery of superconductivity in the Fe-based pnictides has prompted vigorous research activities on complex FeAs-based compounds and related systems [1]. The superconducting iron pnictides have been closely compared with the high-temperature superconducting cuprates. Both have layered structures, short coherence lengths, unusually high upper-critical fields ($H_{c2}$), possible gap-nodes, and possible competitions from magnetic interactions. Despite all efforts, however, the data on these pnictides appears to be rather divergent and confusing. For example, the anisotropy γ, a characteristic of the interlayer coupling, appears to vary widely. Its reported values range from as low as 1.2, *i.e.* almost isotropic, up to 65, *i.e.* similar to that of cuprates [2]. The ability to associate superconductivity with low-dimensionality, *i.e.* the transition temperatures can be significantly raised through altering the inter-layer coupling, depends on an improved understanding of these divergent data. Similar situations exist for the pairing symmetry. Various data have been used to argue that either there are possible gap-nodes [3] or several coexisting gaps [4-7]. Even among the data that prefer a multigap *s*-wave pairing, the coupling strength, $\alpha = 2\Delta/kT$, spreads significantly, ranging from larger than 7 down to close to 1 [4-7]. Such controversy even exists within samples of the same compound. Although the lower critical field, $H_{c1}$, of a $Ba_{0.6}K_{0.4}Fe_2As_2$ single crystal shows clear evidence for the existence of α as small as 1.1 [6], the specific heat of a similar crystal appears to be rather different [8]. Part of the reason for such confusion may be the availability of high quality single crystals. The limited data sets reported thus far have made a systematic interpretation difficult. An especially interesting case is the $Li_{1-x}FeAs$ system, in which superconductivity has been discovered near x = 0 [9-11]. As opposed to other FeAs-based superconducting systems, no static-magnetic-orders have been reported in this system [11-13]. Although lower critical fields on ceramic samples have been reported [13], single crystals large enough to study have become available only recently. Herein we report our work on the magnetic properties of LiFeAs. The lower critical fields, $H_{c1}$, are deduced from both the vortex penetration into single crystals and the diamagnetic moments of a polycrystalline sample. A rather low anisotropy $\gamma \approx 1-2$ is observed, and the T-dependency of the $H_{c1}$ can be better fit with a two-gap model, in good agreement with the reported $H_{c1}$ of $Ba_{0.6}K_{0.4}Fe_2As_2$ [6]. In particular, the observed kink around $T_c/2$ can be understood by invoking a second gap with weaker coupling strength, although our moderate resolution cannot rule out the possibility of either nodal gap or more complicated multi-gap configurations. The result is also in good agreement with our $C_p$ data on similar crystals [14].

Polycrystalline LiFeAs powders were synthesized from high temperature reactions of high purity Li, Fe, and As, as previously reported [10]. The Li deficiency has not been directly measured, but the use of the same preparation procedure and the nearly identical diamagnetic transition lead us to believe that the stoichiometry of the crystals is close to that reported in Ref. 10. Several single crystals with shiny cleavage surfaces and in-layer dimensions of 0.1 mm or larger were chosen and isolated from homogenous bulk samples. The magnetization was measured with a Quantum Design SQUID magnetometer.

A round sheet-like crystal (crystal A) with moderate irregularity, as well as a square crystal (crystal B), were used. The *c* axis is taken as being perpendicular to the cleavage surface, which was verified by Raman spectra of the crystal surfaces. The volume $3.1 \cdot 10^{-5}$ cm$^3$ ($1.1 \cdot 10^{-5}$ cm$^3$) and the demagnetization factor $n_C = H/4\pi M \approx 0.65$ (0.70) along *c* were deduced for crystal A (crystal B) based on the low-field moments M. The observed superconducting transitions, $T_c$, are rather sharp with an onset slightly above 17 K and a transition width of 2-3 K (inset, Fig. 1b).

However, a bulk $T_c$ of 15.3 K, *i.e.* the temperature where 80% of the diamagnetic drop is reached, is used here since the vortex penetration should occur in the weakest sections of the sample.

Extracting the values of lower critical fields from the isothermal low-field M(H) is a rather difficult and sometimes debatable process. Experimentally, the field $H_p$, where the first vortex penetration occurs, is difficult to identify accurately over the smooth M-H observed (Fig. 1a). A procedure, which assumes that the deduced $(M-\chi_{ini}H)^{1/2}$ is zero below $H_p$ but a linear function of H above $H_p$, has been widely accepted for elliptical samples, where $\chi_{ini}$ is the initial susceptibility $\left.\frac{dM}{dH}\right|_{H\to 0}$ [15]. The main idea, *i.e.* a mixing-state layer should be formed on the whole surface outside an untouched Meissner core with an H-independent demagnetization factor, appears to work well for ellipsoids. Using the Bean model, $(H-H_p)$ and $(M-\chi_{ini}H)^{1/2}$ are proportional to the thickness and the effective volume of the penetrated layer, respectively, under such bulk penetration. This bulk penetration assumption, however, may be severely violated for irregular samples with sharp edges. The small sizes and the air-sensitive nature of LiFeAs, unfortunately, forced us to use the crystals as received, *i.e.* with irregular sharp edges. The deduced $(M-\chi_{ini}H)^{1/2}$, for example, is plotted in Fig. 1b for crystal A at 3 K with H//*c*. The $(M-\chi_{ini}H)^{1/2}$ above 150 Oe still appears as a linear function of H, indicating that a bulk penetration against a residual Meissner core is finally reached. The data below 150 Oe, however, deviates significantly from the expected horizontal lines, revealing that severe edge penetration occurs at fields as low as 5 Oe. Modified procedures, therefore, have to be developed to accurately separate the bulk penetration from that of the edge. In addition to such experimental difficulties, the possible Bean-Livingston barrier [16] and the geometric barrier, which is significant only when the demagnetization factor is large, further complicate the data interpretation [17]. The upturns of $H_p$ below $T_c/5$ observed in some cuprates, for example, have been attributed to the hysteretic surface barriers instead of the equilibrium lower critical field [17].

To understand the vortex penetration in non-elliptical samples, the differential susceptibilities $\chi = dM/dH$ (Fig. 2a) are deduced from the data in Fig. 1b. There is no noticeable full-Meissner region where the $\chi$ is an H-independent constant. Instead, the low-field $\chi$ appears to be a linear function of H up to 140 Oe. The diamagnetic moments, therefore, can be empirically expressed as $d+eH+fH^2$, where $d$ represents the possible ferromagnetic background. At higher fields up to 2000 Oe (only that below 300 Oe is shown in Fig. 2a), the only noticeable sharp anomaly is the change of the slope $d\chi/dH$ around 140 Oe. The slope appears to be H-independent again above 140 Oe. Similar situations occur in all cases investigated here. It is interesting to note that such low-field $\chi$ actually measures the volume of the residual Meissner core since the moments of the mixed-state layers should be relatively smaller. The slope $d\chi/dH$, therefore, represents the surface area of the untouched core under the Bean model. The turning point of $d\chi/dH$ consequentially should separate the bulk penetration part from the edge one, in our view. A least-square fitting procedure is used to determine this turning point $H_{p1}$ (Fig. 2a). For further verification, the proposed $(M-\chi_{ini}H)^{1/2}$ vs. H procedure was modified. The residual core is expected to be elliptical in the bulk-penetration stage since such a shape keeps the surrounding vortex lines with minimum curvatures. The "baseline," however, should have an initial moment of $M_0 = d+eH_{p1}+fH_{p1}^2$ and a $\chi = e+2fH_{p1}$ at the bulk-penetration stage (marked as the thick dashed line in Fig. 2a). Another least-square code was developed to fit $(M-M_0)^{1/2}$ to zero and the straight line of bulk penetration below and above the first-penetration field $H_{p2}$, respectively (Fig.

2b). The fitting is excellent; the large deviations below 140 Oe in Fig. 1b disappear, and typical deviations are only a few multiples of the data fluctuations ≈ $10^{-7}$ emu. For both of the fitting procedures, the fitting uncertainty of $H_p$, *i.e.* the range where the root-mean-square (*rms*) increases by a factor of two if all other parameters are fixed, is typically smaller than 5 Oe. The differences between $H_{p1}$ and $H_{p2}$ are usually 5-10 Oe or smaller. The results are also independent of the initial value over a broad range of 20-300 Oe. Unfortunately, the parameters are highly correlated, and the traditional fitting uncertainty likely underestimates the possible deviations. We therefore have to estimate the uncertainty by repeated measurements. The results, fortunately, convince us that the typical uncertainty associated with the penetration fields is limited to within 20 Oe.

It is well known that the local field acting on the surface of a superconducting ellipsoid is $H_{ex}/(1-n)$, where $H_{ex}$ and $n$ are the external field and demagnetization factor, respectively [18]. This is a combined result of the continuity of the tangential H component and the constraint B = H-4πM = 0 inside superconductors. A similar situation is expected for the Meissner core if all edge penetrations have not yet significantly screened the surface field, an assumption verified by our calculations. The "first-penetration" field, therefore, can be deduced.

To identify the field as the lower critical field $H_{c1}$, however, the effects of the possible surface barriers have to be explored. This question has been addressed previously. Several methods have been used to identify the barriers: the asymmetry of M-H loops [15]; the dependency on the field sweep-rate [16]; the dependency on the demagnetization factor [16]; and the differences between the H-increase and H-decease branches [19]. Isothermal M-H loops up to ± 3 T over 4-12 K, therefore, were first measured for crystal A with H//$c$ to verify the possible surface barriers. The M-H loops were roughly symmetric between the H-increase and H-decrease branches, *e.g.* that at 10 K with H//$c$ (inset, Fig 1a). The residual asymmetry, on the order of $10^{-5}$ emu (< 1 emu/cm$^3$), is larger than the experimental fluctuations around $10^{-6}$ emu but in line with the reversible diamagnetic moments observed in ceramic samples (as will be discussed below). Similar situations also occur at other temperatures.

The method of comparing the H-increase and the H-decrease branches, unfortunately, may not be straightforward here with significant irreversible edge penetration. In addition to the symmetry of M-H loops, both the sweep-rate dependency and the dependency on the demagnetization factor have also been checked. The results are independent of the sweep-rate down to $10^{-2}$ Oe/s. The $H_p$ of crystal A at H//$ab$, where the $n_i$ is only 0.17, shows almost the same T dependency. The data for crystal B show comparable results. It is especially interesting to note that the surface quality, a key parameter for the strength of surface barriers [16], should be rather different for the three sequential measurements. The unavoidable air exposure during changing the crystal orientation should affect the possible surface barriers noticeably, as suggested by the surface brightness observed. The consistent T dependency, therefore, suggests that significant surface barriers are unlikely. The deduced $H_p$ is consequentially adopted as the lower critical field (Fig. 3).

To compare the data with various pairing models, the phenomenological procedure reported by A. Carrington and F. Manzano is used [20], *i.e.* taking the normalized superfluid density as:

$$H_{c1}/H_{c1}(0) = \rho_S = 1 + 2\int_{\Delta}^{\infty} dE \frac{\partial f(E)}{\partial E} \cdot \frac{E}{\sqrt{E^2 - \Delta(t)^2}} = 1 + 2\int_0^{\infty} \frac{e^{\sqrt{\varepsilon^2 + \Delta(t)^2}} d\varepsilon}{t(e^{\sqrt{\varepsilon^2 + \Delta(t)^2}} + 1)^2} \quad (1)$$

with the approximation of $\Delta(t) = \Delta_0 \tanh\{1.82[1.018(1/t-1)]^{0.51}\}$, where $\varepsilon$ and $t$ are the energy of normal-electrons and the deduced temperature $t = T/T_c$, respectively. The normalized $H_{c1}$ with two gaps, $\Delta_1$ and $\Delta_2$, at a mixing ratio of r, is then a simple sum $r \cdot \rho_S(\Delta_1) + (1-r) \cdot \rho_S(\Delta_2)$. The data observed can be well fit by a two-gap *s*-wave pairing (solid line, Fig. 3) with $\Delta_1 = (2.7\pm0.8)kT_c = 3.3$ meV, $\Delta_2 = (0.5\pm0.2)kT_c = 0.6$ meV, and $r = 0.5\pm0.2$. The extrapolated $H_{c1}(0)$ will be ≈ 380 Oe. A single *s*-wave gap, on the other hand, leads to a rather different trend: it largely misses the kink around 4-9 K (Fig. 3). The temperature range and the moderate resolution here, unfortunately, would not be able to exclude the possible gap nodes. However, a single-gap *d*-wave pairing, which leads to a quasilinear T dependency of $H_{c1}$ in cuprates [17], cannot reproduce the observed kink. A multi-gap pairing should be a likely scenario. It is interesting to note that a recent ARPES work on LiFeAs suggests that the gap-widths over different Fermi surface pockets are noticeably different, *i.e.* being 1.5 meV and 2.5 meV over the hole-like and electron-like parts, respectively [21]. Although the extracted values are only in rough agreement with the 0.6 meV and 3.3 meV reported here, such multigap feature seems to be rather natural with the multi conducting bands of the FeAs-based superconductors. It should be pointed out that although our moderate resolution may not allow us to distinguish two closely located gaps, *e.g.* at $3.5kT_c$ and $2.5kT_c$, the kink around 0.5 $T_c$ can hardly be accommodated without a small gap. Similar features have also been reported in the $H_{c1}$/penetration depth of other pnictides [6, 22]. It is especially interesting to note that such a small gap has also been in line with our specific heat data on a similar LiFeAs crystal [14].

Both the gap values and the mixing ratio so deduced are in rough agreement with those reported for $Ba_{0.6}K_{0.4}FeAs$ single crystals, *i.e.* $3kT_c$, $0.7kT_c$, and 0.3, respectively [6]. The lower gap in both cases is smaller but significantly affects the zero-temperature superfluid density. The results, however, are rather different from those of $(Ba,K)Fe_2As_2$ deduced from ARPES (two gaps ≈ $3.5kT_c$ and $2.5kT_c$, respectively) [4], from tunneling experiments (≈ $4kT_c$ and $1.3kT_c$) [5], or even from specific-heat data (single gap around $2kT_c$) [8]. We do not yet have a good understanding for such discrepancies. However, it is interesting to note a trend that the contribution of the narrower gap seems to affect $H_{c1}$ more prominently. The mixing ratio $r = 0.7$ extracted from the $C_p$ of similar LiFeAs crystals are also noticeably larger [14], and we noticed a similar trend in our investigation of $(Ba,K)Fe_2As_2$.

The anisotropy of the lower critical fields is shown in Fig. 4. The average ratio $H_{c1}(H//c)/H_{c1}(H//a,b)$ is only 1.2±0.2 between 4 and 12 K for crystal A and 1.3±0.2 for crystal B. Although it is broadly accepted that the FeAs-based superconductors possess a moderate anisotropy around 3-5, the reported data appear to vary significantly [2,23]. Anisotropy as low as 1.2, in particular, has been reported on $(Ba,K)Fe_2As_2$ single crystals [2,24]. Our LiFeAs crystals belong to this category.

To further deduce the superconducting parameters, the magnetization of a polycrystalline sample was analyzed based on the modified London model [25]. It is interesting to note that the macroscopic magnetizations have rarely been analyzed in the FeAs-based superconductors. The significant, but poorly understood, magnetic background is the main reason, which often dominates the data under high fields. We have significantly suppressed the background by proper after-synthesis anneals. While the residual background is still large, it is insensitive to the temperature above $T_c$ and with negligible hysteresis (inset, Fig. 5). This enables us to extract important parameters and further verify the possible surface barriers.

The raw M-H data are shown in the inset in Fig. 5. The magnetic contribution, *e.g.* the M at 18 K (solid line), is similar to that of soft ferromagnets with negligible hysteresis. The magnetizations further demonstrate a rather weak T dependency, *i.e.* within a few percent from 18 to 25 K. The magnetization at 18 K, therefore, was used as the background. It is interesting to note again that nearly all magnetizations at the field-decrease branches below $T_c$ are noticeably lower than the moments at 18 K above 1 T. The Bean-Livingston barriers, therefore, are unlikely to affect the moments significantly. The average moments between the H-increase and the H-decrease branches were taken as the reversible diamagnetic moment. This is supported by the observation that these moments vary with the field logarithmically as expected (Fig. 5). The modified London model, therefore, is applied. The reversible moment in the model will be $M = -\dfrac{\alpha \phi_0}{32\pi^2 \lambda^2} \ln(\dfrac{\beta H_{c2}}{H})$, where $\phi_0$, $\lambda$, $H_{c2}$, $\alpha \approx 0.77$, and $\beta \approx 1.44$ are the flux quantum, penetration depth, the upper critical field, and two numerical factors, respectively. Consequentially, the slope will be $\dfrac{\partial M}{\partial \ln H} = \dfrac{\alpha \phi_0}{32\pi^2 \lambda^2}$ and the intercept $\approx \beta H_{c2}$. The polycrystalline nature of the sample should not invoke significant modification here based on the low anisotropy observed. The lower critical field $H_{c1}$ and the Ginsburg-Landau parameter $\kappa$ were then regressively calculated. A T-independent $\kappa \approx 25$ is then obtained. The $H_{c1}$, so obtained, is in reasonable agreement with those deduced above (Fig. 3). The penetration depth of two ceramic LiFeAs samples was previously reported [13]. The corresponding lower critical fields $\leq 200$ Oe for $\kappa < 200$, however, are significantly lower than the values deduced here. Exact reasons for such disagreement are as yet unknown, but the differences in doping level and defect density might play a role.

In summary, the lower critical fields of LiFeAs have been deduced from both the vortex penetration of a single crystal and the reversible magnetization of a polycrystalline sample. The compound seems to be almost isotropic, and can be fit as a two-gap superconductor with the superfluid density strongly affected by the smaller gap.

**Acknowledgments**

We thank Dr. V. Hadjiev for help in the Raman measurements of the crystals. The work in Houston is supported in part by the U. S. Air Force Office of Scientific Research, the T. L. L. Temple Foundation, the John J. and Rebecca Moores Endowment, and the State of Texas through the Texas Center for Superconductivity at the University of Houston; and at Lawrence Berkeley Laboratory by the Director, Office of Science, Office of Basic Energy Sciences, Division of Materials Sciences and Engineering of the U.S. Department of Energy under Contract No. DE-AC03-76SF00098. A.M.G., B.L., and Z.T. acknowledge support from the NSF (CHE-0616805) and the R.A. Welch Foundation.

**Figure Captions**

Fig. 1:  a) The virgin isothermal M(H) under zero-field-cool condition of a sheet-like single crystal (crystal A) at (from bottom up) 2, 3, 4, 5, 6, 7, and 8 K, respectively. Inset: The M-H loop at 10 K. The sample volume is calculated using the low-field susceptibilities. b) The calculated $\sqrt{M - \chi_{H=0} H}$ of crystal A along the *c*-axis based on the $\chi_{H=0}$ calculated below 10 Oe. The straight line is the model expectation. Inset: The zero-field-cool magnetization at 10 Oe with H//c.

Fig. 2:  a) The observed dM/dH vs. H for crystal A at 3 K and H//*c*. The symbols are the data, and the lines are the linear fits at lower and higher fields, respectively. The thick dashed line corresponds to the "baseline" $M_0$ used (see text). b) The deduced sign(M-$M_0$)·| M-$M_0$ |$^{1/2}$ (circles). The line is the best fit.

Fig. 3:  The lower critical fields of LiFeAs crystal A at H//*c*. Solid triangles: deduced $H_{p2}$; open triangles: deduced $H_{p1}$; open squares: deduced from the reversible magnetization of the ceramic; solid line: two-BCS-gap fit; dashed line: one-gap fit.

Fig. 4:  The anisotropy of $H_{c1}$. Open triangles: crystal A. Solid squares: crystal B.

Fig. 5:  The average moment of the H-increase and the H-decrease branches along the isothermal M-H loops after subtracting those at 18 K > $T_C$. Circles: 10 K; triangles: 12 K; and squares: 14 K. Inset: The raw M-H loops. The line is the moment at 18 K.

Fig. 1

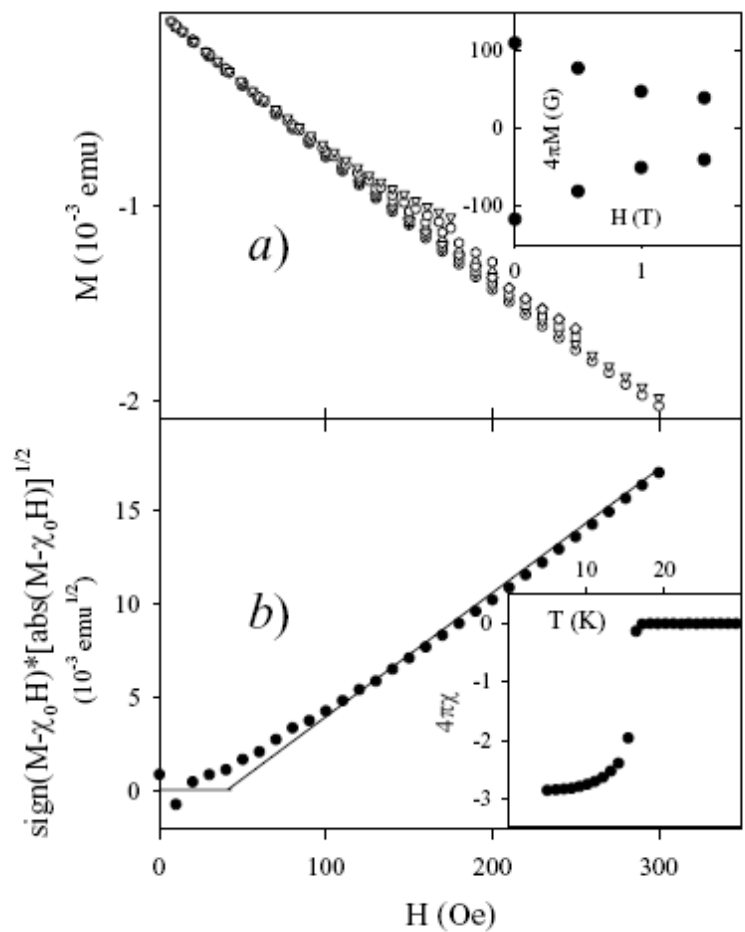

Fig. 2

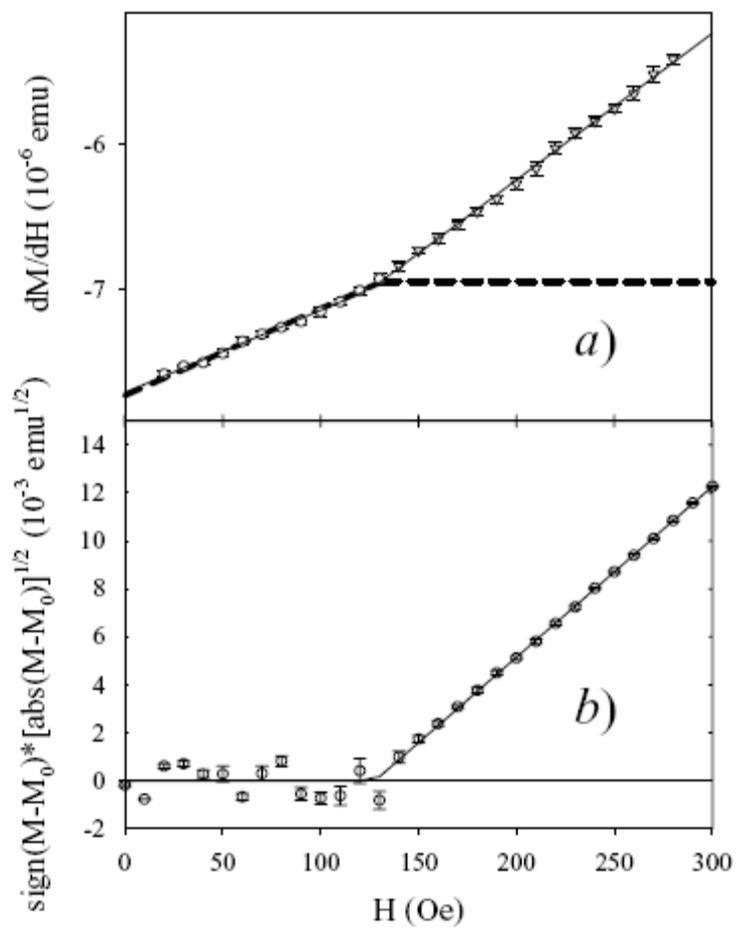

Fig. 3

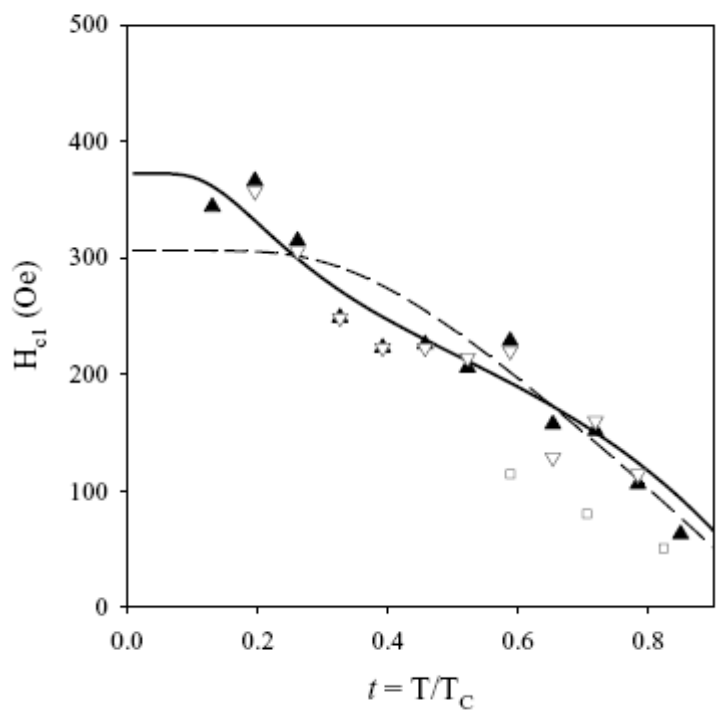

Fig. 4

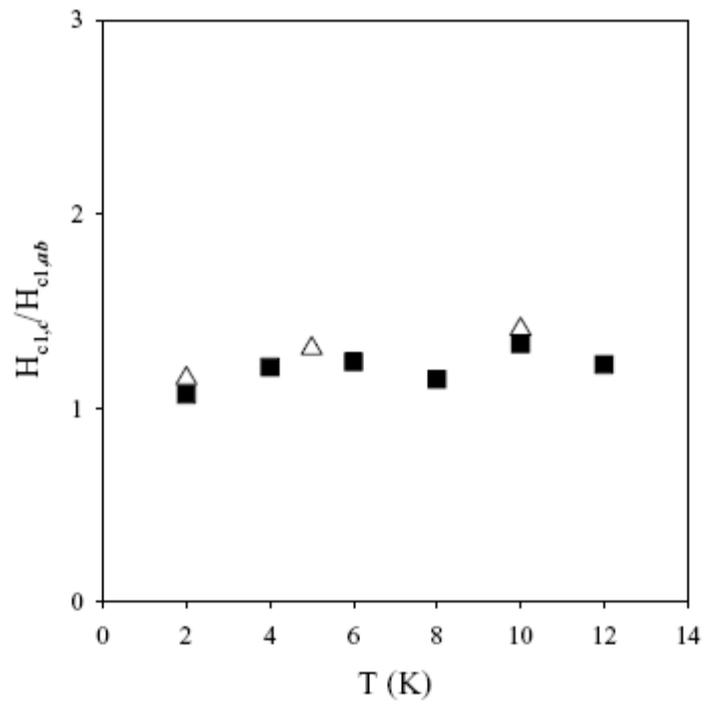

Fig. 5

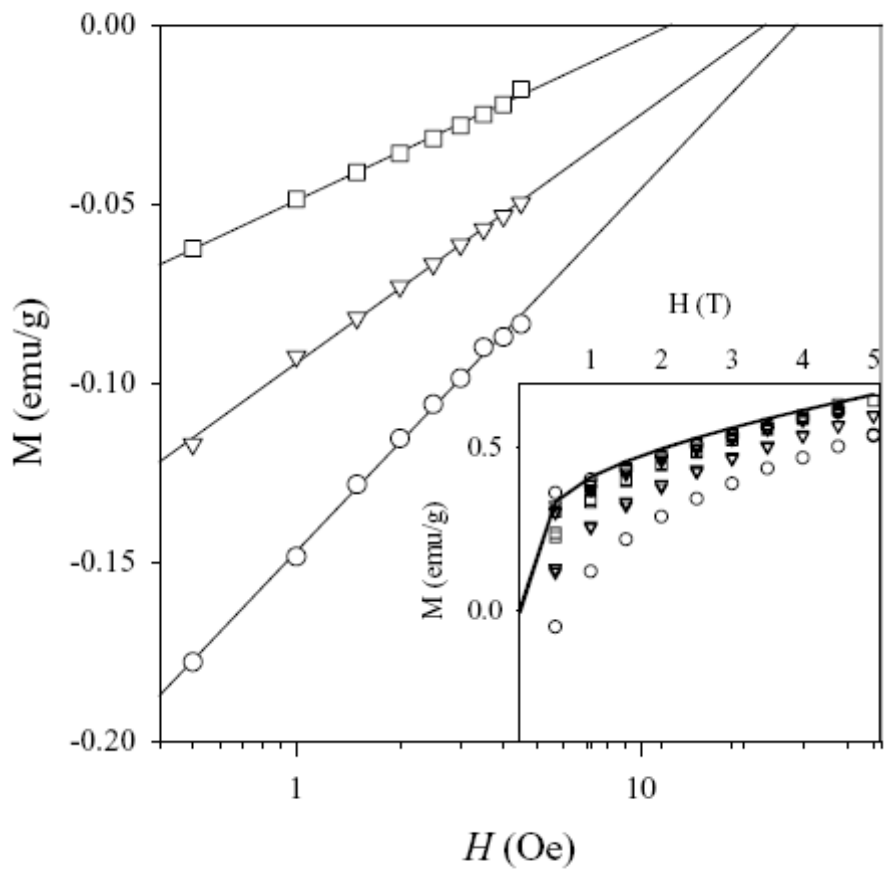